\documentclass[iicol,sn-mathphys-num]{sn-jnl}

\usepackage{graphicx}%
\usepackage{multirow}%
\usepackage{amsmath,amssymb,amsfonts}%
\usepackage{amsthm}%
\usepackage{mathrsfs}%
\usepackage[title]{appendix}%
\usepackage{xcolor}%
\usepackage{textcomp}%
\usepackage{manyfoot}%
\usepackage{booktabs}%
\usepackage{algorithm}%
\usepackage{algorithmicx}%
\usepackage{algpseudocode}%
\usepackage{listings}%

\theoremstyle{thmstyleone}%

\theoremstyle{thmstyletwo}%

\theoremstyle{thmstylethree}%

\begin{document}

\title[Article Title]{Study of the Log-third-difference method for the computation of even-odd staggering in fission yields}

\author[1]{\fnm{Bel\'en} \sur{Montenegro Vi\~nas}}\email{belenmvinhas@gmail.com}
\author*[2]{\fnm{Manuel} \sur{Caama\~no}}\email{manuel.fresco@usc.es}

\affil[1]{\orgname{Universidade de Santiago de Compostela}, \city{Santiago de Compostela}, \postcode{15782}, \country{Spain}}
\affil[2]{\orgdiv{IGFAE}, \orgname{Universidade de Santiago de Compostela}, \city{Santiago de Compostela}, \postcode{15782}, \country{Spain}}

\abstract{The even-odd staggering in fission fragments yield distributions is an important observable to address the role of intrinsic excitations in the fission process. However, its computation as a function of the fragment split is an ill-posed problem that is usually solved through certain assumptions. Arguably, the most widely used method for its computation is the Log-third-difference method, which assumes an underlying local Gaussian distribution. Experimental data of the even-odd effect show sharp features and peaks that are a challenge for the Log-third difference method (or any method for that matter). This paper presents a detailed study of the performance of the Log-third difference method and describes three strategies for improving the precision and accuracy around these sharp features. The best results are obtained with an iterative application of the method and with the use of a multi-Gaussian fit. In regions with a smooth behaviour, the direct application of the Log-third-difference method remains the best option.}

\maketitle

\section{Introduction}\label{sec1}

Nuclear fission is one of the most complex decay process in Nuclear Physics: a heavy atomic nucleus rearranges its nucleons and undergoes extreme deformations to the point of splitting into two smaller fragments, while releasing a considerable amount of energy~\cite{meiN39}. The process is driven by both macroscopic and microscopic properties of nuclear matter, and the output is a consequence of their competition~\cite{benJPG20}.

In particular, the fragment size distribution is usually characterised with several {\it modes}, which correspond to specific partitions favoured by microscopic and macroscopic properties~\cite{broPR90}. Each mode is regarded as a set of events with mass, neutron or proton partitions that fluctuate around a well-defined value. Phenomenologically, these are described as Gaussian distributions centred at specific values of mass (A), neutron (N) or proton (Z) number and with amplitudes and widths usually determined by fitting experimental data. Although their nature and origin, and even their existence, are still a matter of study~\cite{broPR90,wilPRC76,bocNPA08,scaN18}, fission modes are widely used in fission studies.

As soon as experiments were able to determine the fragment Z, the corresponding yield distributions from low-energy fission revealed a persistent pattern: in even-Z fissioning systems, fragments with an even proton were more produced than odd-Z fragments, which is translated to an even-odd staggering in the fragment-Z yields~\cite{amiPRC75}. In order to quantify this property, the amplitude of the total, or global, even-odd effect $\delta_z^{tot}$ was defined as the normalised difference between the total even-Z and odd-Z fragment yields:
\begin{equation}
\delta_z^{tot}=\frac{\sum_{{\rm even}~{\rm Z}} Y({\rm Z})-\sum_{{\rm odd}~{\rm Z}} Y({\rm Z})}{\sum Y({\rm Z})}.
\label{eq_delta}
\end{equation}

The amplitude of the even-odd effect was found to depend on the fissility of the fissioning system and on the initial excitation energy. These correlations were explained by identifying intrinsic excitations and dissipation mechanisms as the origin of the proton even-odd effect~\cite{nifZPA82,gon91,pomNPA93,perNPA97,rejNPA00}. A similar effect is expected also in the neutron number, but the smaller neutron binding energies and the fragment neutron evaporation makes its experimental study more challenging.

The experimental fragment-Z yield distributions also suggested a certain evolution of the effect as a function of the split. In order to study this evolution, different prescriptions to evaluate the amplitude of the even-odd effect as a function of Z were proposed~\cite{gonNIM,olmEPJA15}. Among them, the most widely used and, arguably, the most accurate is the Log-third-difference formula suggested by Tracy~et~al. in~\cite{tracy}. This formula computes a weighted average of $\delta_z$ with the yields of four consecutive fragment-Z, assuming that the yields without staggering follow a Gaussian behaviour:

\begin{equation}
\begin{aligned}
\delta_z^{\rm T}({\rm Z+1.5})&= \frac{(-1)^{{\rm Z}}}{8}\Big( \ln{Y({\rm Z})}-\ln{Y({\rm Z+3})} \\
& + 3\big[ \ln{Y({\rm Z+2})}-\ln{Y({\rm Z+1})} \big]\Big).
\end{aligned}
\label{eq_tracy}
\end{equation}

The possibility of studying the {\it local} even-odd effect $\delta_z({\rm Z})$ revealed new properties~\cite{steinNPA,caaJPG} and helped to understand the mechanism that produces it, and its relation with the dynamics and energetics of the fission process~\cite{schPRL,jurJPG15}. Among these properties, the most relevant ones were the increase of the amplitude of the local $\delta_z({\rm Z})$ with the asymmetry of the split and the existence of a negative $\delta_z({\rm Z})$ in the heavy fragments of odd-Z fissioning systems.

However, in recent studies, the evolution of the local $\delta_z({\rm Z})$ suggests dependences on specific proton numbers and on the initial excitation energy that manifest in abrupt changes or sharp features, such as narrow peaks, that would require an improved version of the classical formula in order to perform an accurate characterisation. In this paper, we review the limitations of the Log-third difference prescription and explore methods to reduce their impact around these sharp features.\footnote{It is important to note that the properties and methods are discussed in a general basis but better results may be attained by tailoring their application for each case individually.}

\section{The Log-third-difference formula}
\label{sec_tracy}

As already mentioned, Eq.~\ref{eq_tracy} computes a weighted average of $\delta_z$ for a set of four consecutive yields, providing they follow an underlying Gaussian behaviour. More specifically, proton yields can be described with a smooth behaviour modulated by a certain even-odd effect:
\begin{equation}
Y({\rm Z})=Y^{s}({\rm Z})\left[1+(-1)^Z\delta_z({\rm Z})\right]. 
\end{equation}
In this way, we can rearrange the terms of Eq.~\ref{eq_tracy} as
\begin{equation}
\delta_z^{\rm T}({\rm Z+1.5})= (-1)^{{\rm Z}}\big[\Delta G + \langle\langle\delta_z\rangle\rangle\big],
\label{eq_tracycomps}
\end{equation}
with
\begin{equation}
\Delta G=\frac{1}{8}\left[ \ln\frac{Y^s({\rm Z_0})}{Y^s({\rm Z_3})}+3\ln\frac{Y^s({\rm Z_2})}{Y^s({\rm Z_1})}\right]
\end{equation}
and
\begin{equation}
\begin{aligned}
\langle\langle\delta_z\rangle\rangle&=\frac{1}{8}\Big[\ln\Big(1+(-1)^{Z}\delta_z({\rm Z})\Big)\\
&-\ln\Big(1-(-1)^{Z}\delta_z({\rm Z_3})\Big)\\
&+3\ln\Big(1+(-1)^{Z}\delta_z({\rm Z_2})\Big)\\
&-3\ln\Big(1-(-1)^{Z}\delta_z({\rm Z_1})\Big)\Big],
\end{aligned}
\label{eq_avdelta}
\end{equation}
with \mbox{${\rm Z}_i={\rm Z}+i$}. 

The $\Delta G$ term is a measure of how much the underlying $Y^s({\rm Z})$ deviate from a Gaussian behaviour, being \mbox{$\Delta G=0$} when they follow a Gaussian function. The $\langle\langle\delta_z\rangle\rangle$ term is an approximate weighted average of the $\delta_z({\rm Z})$ of the four-Z region.

Considering that \mbox{$\ln\big(1+(-1)^{Z}\delta_z\big)\approx (-1)^{Z}\delta_z$} for small values of $\delta_z$, if the smooth yields $Y^s({\rm Z})$ follow a Gaussian function and $\delta_z({\rm Z})$ is constant, Eq.~\ref{eq_tracy} results in
\begin{equation}
\delta_z^{\rm T}({\rm Z_{1.5}})= \frac{(-1)^{{Z}}}{2}\Bigg(\ln\frac{1+(-1)^{Z}\delta_z}{1-(-1)^{Z}\delta_z}\Bigg)\approx (-1)^{Z}\delta_z,
\label{eq_G1}
\end{equation}
although, the approximation can be avoided with a corrected version of $\delta_z^{\rm T}$:
\begin{equation}
\delta_z^{\rm T*}=\frac{e^{2\delta_z^{\rm T}}-1}{e^{2\delta_z^{\rm T}}+1}
\label{eq_tracycorr}
\end{equation}

This brief analysis already hints at some limitations in the formula: the estimated $\delta_z^{\rm T*}$ will deviate from $\delta_z$ when the underlying behaviour is not Gaussian-like and/or when $\delta_z$ changes rapidly within the four-Z set. Another issue is that the result of the formula corresponds to \mbox{Z+1.5}, a non-integer value of fragment Z, which may be a nuisance when correlating the behaviour of $\delta_z$ with particular Z numbers~\cite{ramPRC23}. These issues, and suggested methods to reduce them, are explored with a set of simulations that replicate the typical behaviour of fission fragment yields and $\delta_z$, and detailed in the next section.

\section{Simulation of fragment yields and even-odd effect}
\label{sec_sim}
The simulation used in this work generates an underlying smooth yield distribution, which is then modulated with a $\delta_z({\rm Z})$ generated independently.

The first step is to generate the atomic number of the fissioning system ${\rm Z_{FS}}$, which is randomly selected between 84 and~101. The fragments smooth yield distribution is produced with the sum of two asymmetric and one symmetric mode, described as Gaussian distributions. In the case of the symmetric mode, the distribution is centred at ${\rm Z_{FS}}/2$, with a standard deviation between 4 and~5 units of Z. In the case of the asymmetric modes, both are the sum of two distributions corresponding to the heavy and light fragment, respectively. Following the experimental systematics in this region, the  distributions of the heavy fragment of each mode are centred between \mbox{${\rm Z_{I}}=50$} and~54, and between \mbox{${\rm Z_{II}}=54$} and~58, respectively,\footnote{Loosely, they would correspond to the classical Standard I and II modes.} while their light fragment counterparts are centred at \mbox{${\rm Z_{FS}}-{\rm Z_{I}}$} and \mbox{${\rm Z_{FS}}-{\rm Z_{II}}$}. The standard deviation of each mode varies between 2.5 and~3.5 units of Z. The relative amplitude of each mode is randomly selected.

\begin{figure}[t]
 \includegraphics[width=\columnwidth]{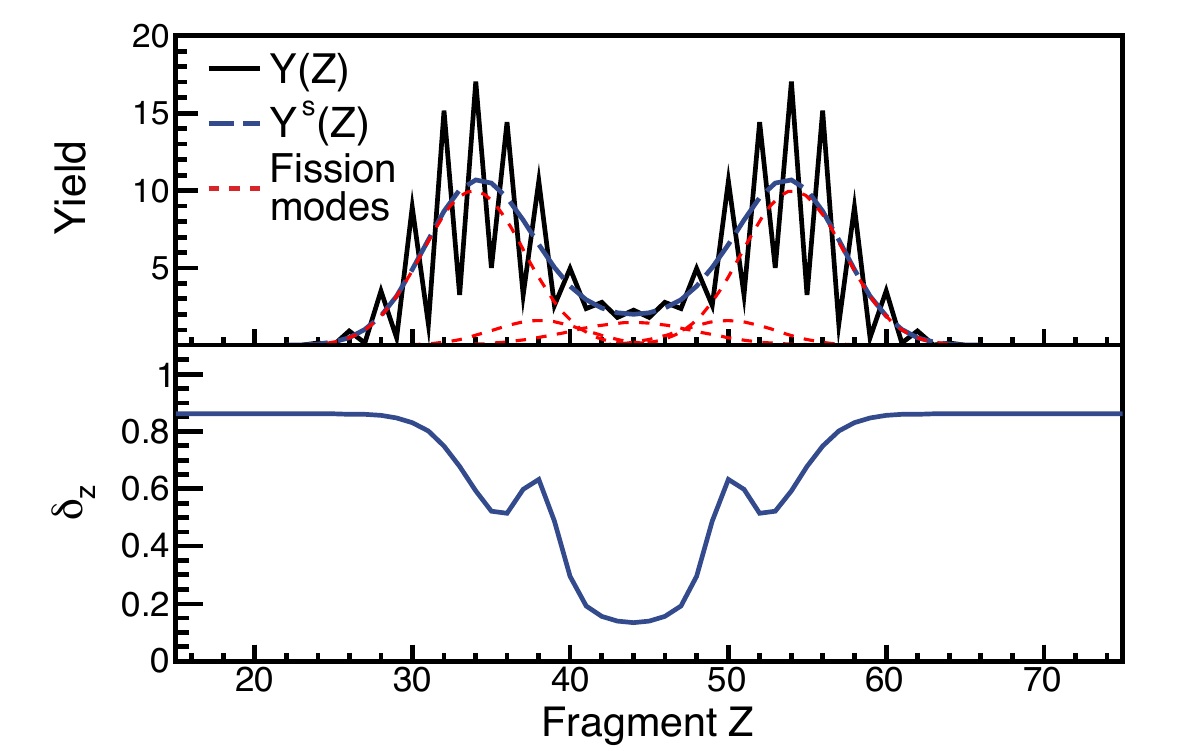}
 \includegraphics[width=\columnwidth]{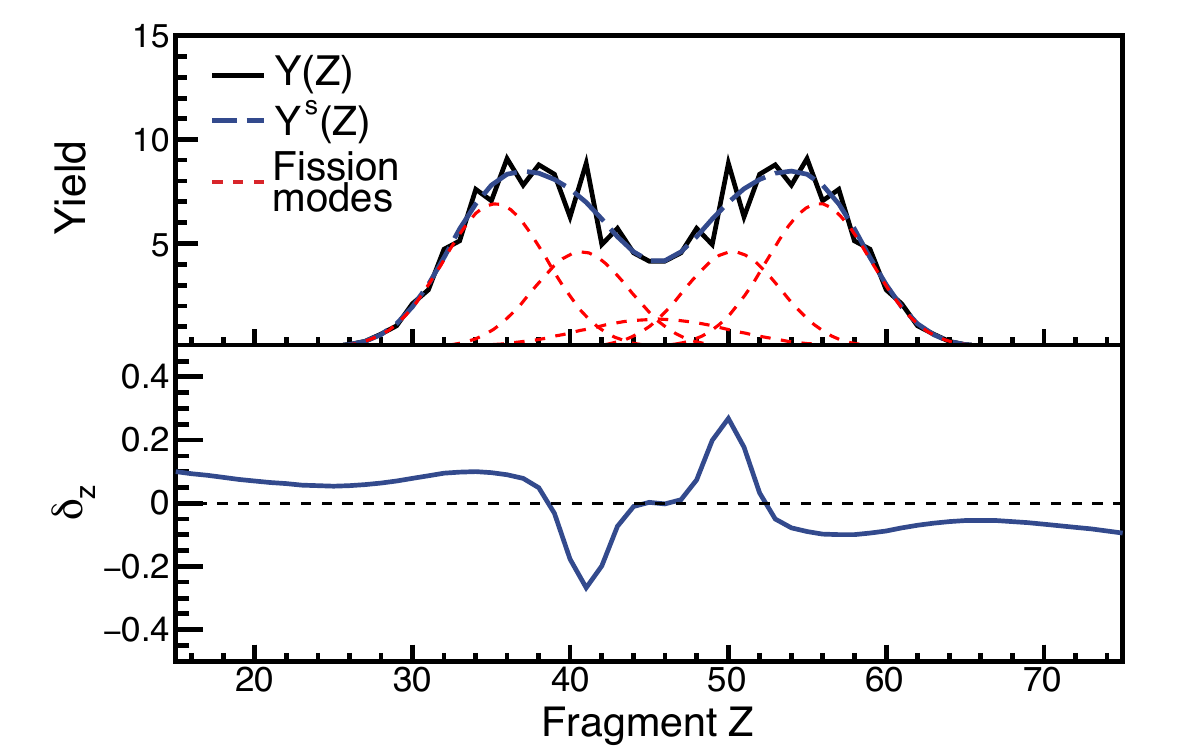}
  \caption{The upper half of each panel shows simulated fragment yields $Y({\rm Z})$ (black line), the smooth underlying distributions $Y^s({\rm Z})$ (long-dashed blue line), and the fission modes (short-dashed red lines). The lower half shows the simulated $\delta_z({\rm Z})$ (blue line). The top panel corresponds to an even-\mbox{${\rm Z_{FS}}=88$} fissioning system, while the bottom panel corresponds to an odd-\mbox{${\rm Z_{FS}}=91$} fissioning system.}
\label{fig1}
\end{figure}

The behaviour of $\delta_z({\rm Z})$ is simulated with the sum of three functions: a wide parabolic shape centred at ${\rm Z_{FS}}/2$ that increases with the asymmetry of the fragments, a relatively narrow Gaussian centred at \mbox{${\rm Z}=50$} with a standard deviation between 0.5 and 1.5 units of Z, and a wider Gaussian function centred between \mbox{${\rm Z}=54$} and~58, and a standard deviation between 2 and~6 units of Z. 

As in the case of the fission yield modes, the Gaussian functions have their counterparts in the light fragment side centred at the complementary fragment Z. If the fissioning system has an even ${\rm Z_{FS}}$, all components, and thus the behaviour of $\delta_z({\rm Z})$, are symmetric with respect to ${\rm Z_{FS}}/2$. For odd-${\rm Z_{FS}}$ systems, $\delta_z({\rm Z})$ is necessarily antisymmetric with respect to ${\rm Z_{FS}}/2$. However, while the parabolic function and the wider Gaussian are negative in the heavy fragment side, the narrow peak centred at \mbox{${\rm Z}=50$} is positive. This behaviour echoes the experimental data from~\cite{steinNPA}, where the amplitude of $\delta_z({\rm Z})$ was found to increase with the asymmetry of the split and negative values were observed in the heavy fragments for \mbox{odd-${\rm Z_{FS}}$}; it also follows the experimental data from~\cite{ramPRC23}, where the strength of the spherical closed shell at \mbox{${\rm Z}=50$} seems to overcome the influence of the level density on the energy flux between pre-fragments. Figure~\ref{fig1} shows an example of randomly generated yields and $\delta_z({\rm Z})$ for even- and odd-${\rm Z_{FS}}$ fissioning systems.

Throughout this paper, the performance assessment of the methods is done with parameters of probability distributions, such as average and width. However, the actual shapes of these distributions are governed by the homogeneous, and somehow arbitrary, distributions of the underlying parameters corresponding to the random generation of yields and even-odd effects. Therefore, instead of using the standard deviation, the width $w(x)$ of a frequency distribution $f(x)$ is given as the limits around the mean value $\bar{x}$ that contain 95\% of the cases:
\begin{equation}
\int_{\bar{x}-\mathbf{w(\mathit{x})}}^{\bar{x}+\mathbf{w(\mathit{x})}}f(x){\rm d}x=0.95\int_{-\infty}^{\infty}f(x){\rm d}x.
\label{eq_w}
\end{equation}

In addition, the error of any estimation of the amplitude of $\delta_z$ done with any method is defined as the difference between the estimation $\delta_z^{est}$ and the real value:\footnote{We choose to use the absolute difference instead of the relative one because there is little correlation between the difference and the real value, as it is shown, for instance, in Eq.~\ref{eq_difG0}}
\begin{equation}
\varepsilon^{est}=\delta_z^{est}-\delta_z.
\label{eq_err}
\end{equation}

\section{Performance and limitations of the Log-third-difference formula}

\begin{figure}[t]
 \includegraphics[width=\columnwidth]{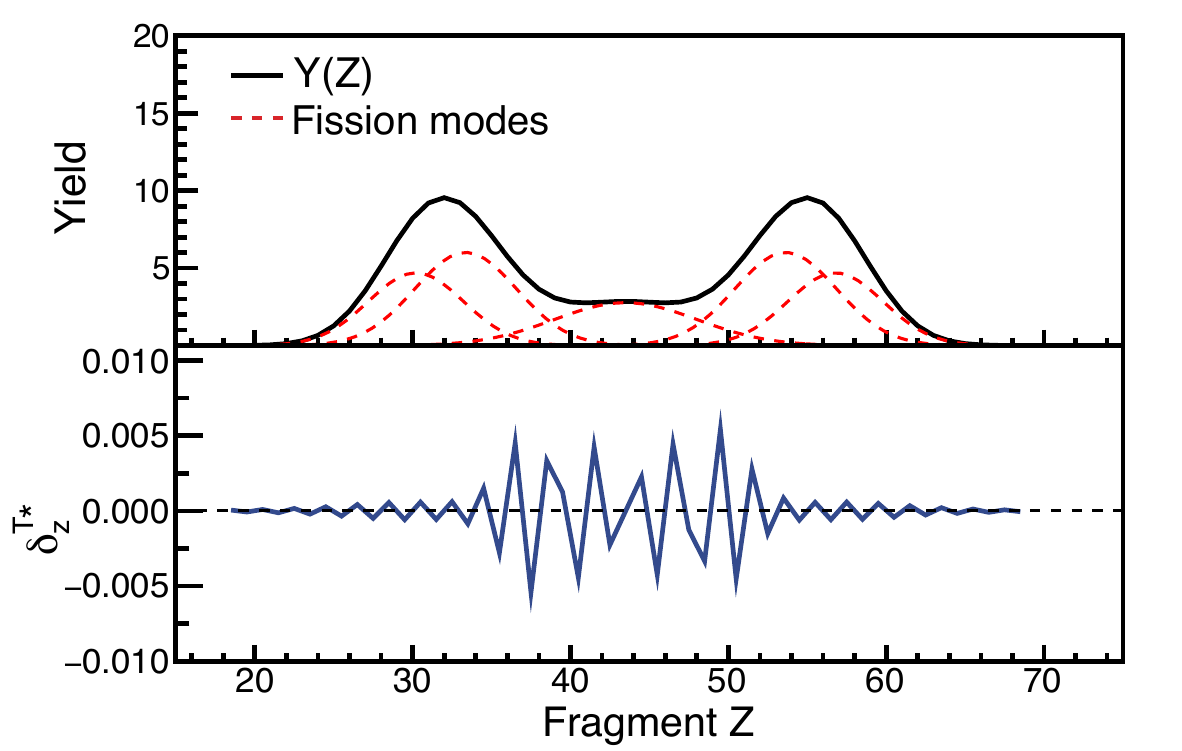}
  \caption{The upper half shows simulated fragment yields $Y({\rm Z})$ (black line) and the fission modes (short-dashed red lines) for a \mbox{${\rm Z_{FS}}=87$} fissioning system with \mbox{$\delta_z=0$}. The lower half shows $\delta_z^{\rm T*}({\rm Z})$ evaluated with Eq.~\ref{eq_tracycorr} (see text).}
\label{fig2}
\end{figure}

\subsection{The case of $\delta_z=0$}
\label{sec_d0}

According to Eq.~\ref{eq_tracycomps}, in the absence of any even-odd effect \mbox{($\delta_z({\rm Z})=0$)}, the Log-third-difference formula is effectively a probe of the Gaussian character of the distribution. In a typical fragment distribution, the non-Gaussian behaviour is mainly ruled by the superposition of modes, being the region around symmetry particularly affected. Figure~\ref{fig2} shows a typical example. 

The resulting $\delta_z^{\rm T*}$ displays strong oscillations with a period of almost one unit of Z. This is due to the role of $\Delta G$ in Eq.~\ref{eq_tracycomps}: while in any particular region, the deviation from a pure Gaussian behaviour may change gradually as a function of Z, $\delta_z^{\rm T*}$ modulates $\Delta G$ with $(-1)^Z$, hence the observed oscillations.

\begin{figure}[t]
 \includegraphics[width=\columnwidth]{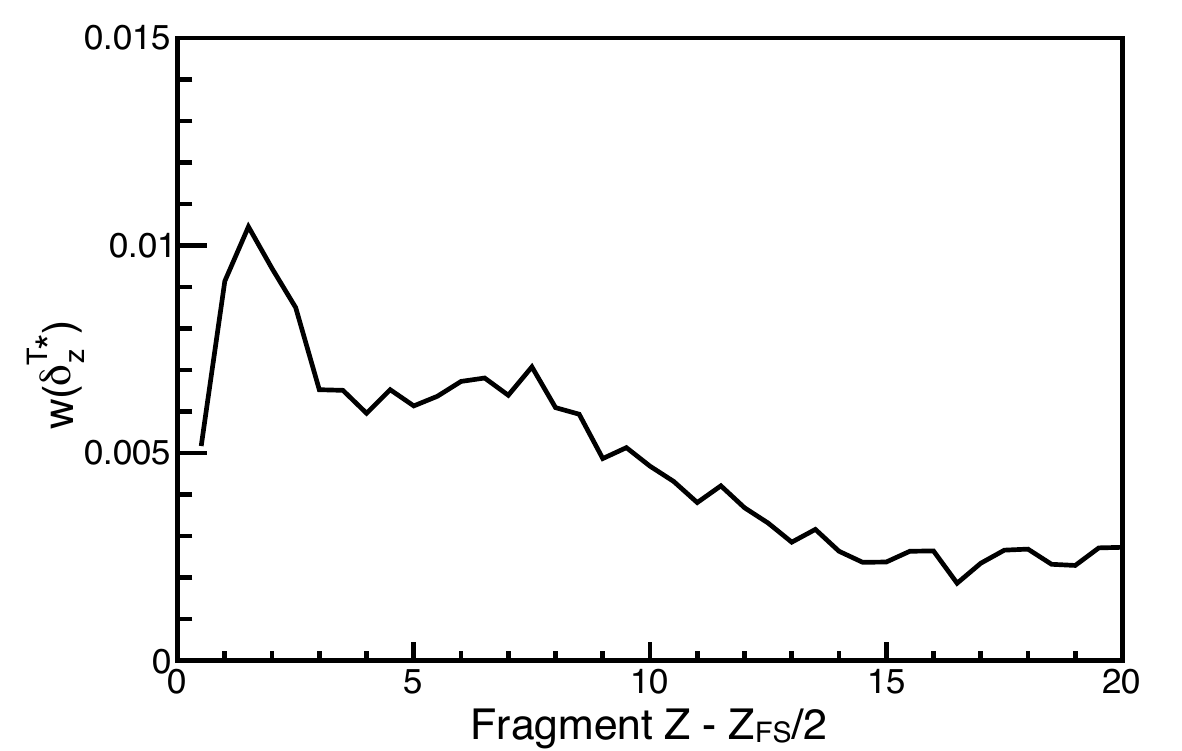}
  \caption{The figure shows $w(\delta_z^{\rm T*})$ for simulated fragment distributions with with \mbox{$\delta_z=0$}.}
\label{fig3}
\end{figure}

However, the amplitude of $\Delta G$ is relatively small, below $\sim$0.01 at symmetry in 95\% of the cases, as it is shown in Fig.~\ref{fig3}.\footnote{With the exception of odd-Z$_{\rm FS}$ systems, where \mbox{$\delta_z^{\rm T*}=0$} at symmetry by construction.} In the same figure, it is possible to recognise different regions defined by the mixture of modes in $\Delta G$: the aforementioned region of symmetry, where $\Delta G$ reaches~0.01; a region stretching to around ten units of Z away from the symmetry with $\Delta G$ above 0.005, where a mixture of SI and SII modes typically appear; and the tail of the distributions, where the contribution from different modes is expected to be small and $\Delta G$ falls below $\sim$0.003.

It is possible to demonstrate that statistical fluctuations linearly increase the uncertainty of the corrected Log-third-difference formula as (see~\ref{app0}):
\begin{equation}
\sigma_{\delta_z^{\rm T*}}\approx\sqrt{\frac{5}{16}}\sigma_{R}\left(1-\delta_z^2\right)
\label{eq_stat}
\end{equation}
where $\sigma_{\delta_z^{\rm T*}}$ is the uncertainty of $\delta_z^{\rm T*}$ and $\sigma_{R}$ is the statistical uncertainty relative to the yields.

In order to be sensitive to an amplitude of $\Delta G\sim 0.01$, and according to Eq.~\ref{eq_stat}, $\sigma_{R}$ should be below 0.018, which corresponds to an experimental yield of \mbox{$Y({\rm Z})\sim 3\cdot 10^3$} events. In fact, this points to a limit for the sensitivity of $\delta_z^{\rm T*}$: absolute values measured below $\sqrt{2\sigma_{\delta_z^{\rm T*}}^2+w(\delta_z^{\rm T*})^2}$ have a reduced significance and may be compatible with \mbox{$\delta_z=0$}. A possible way to reduce this limit for a particular fragment distribution would be to study the behaviour of $\Delta G$ in that distribution with fits as those discussed in Section~\ref{gaus_sec}

\begin{figure}[t]
 \includegraphics[width=\columnwidth]{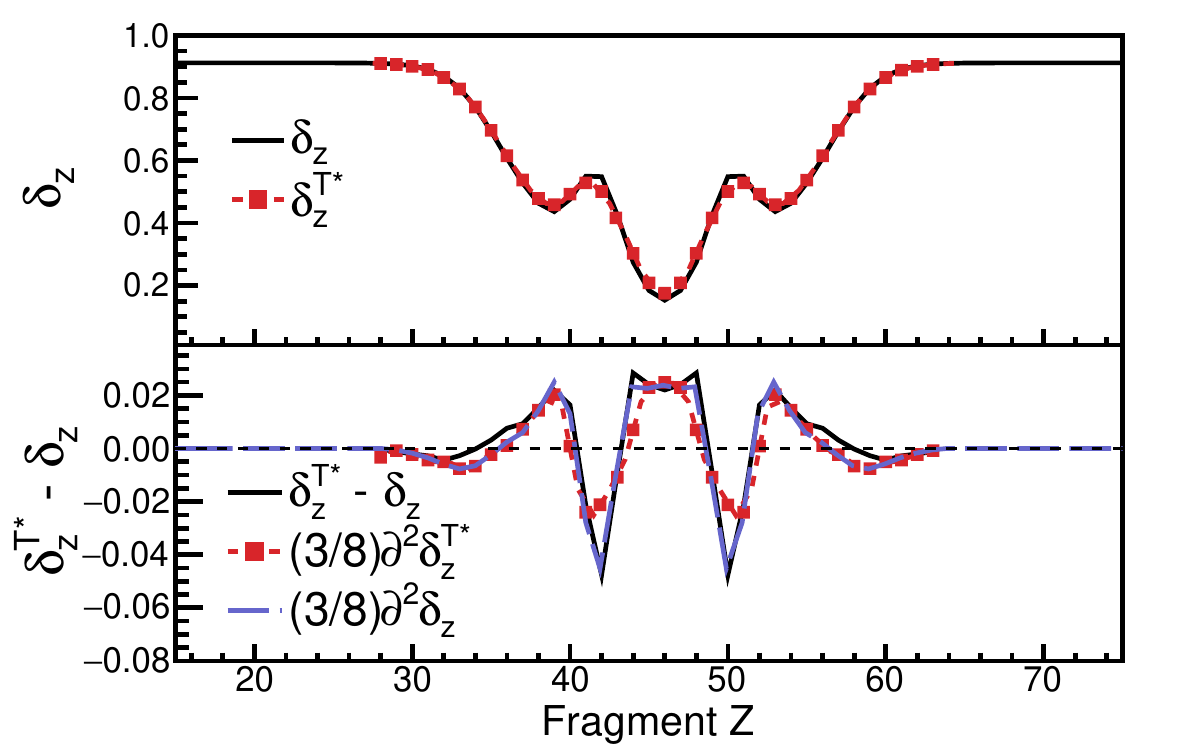}
  \caption{The upper half of the figure shows a simulated $\delta_z$ (solid black line) and the resulting $\delta_z^{\rm T*}$ (dashed red line) for \mbox{Z$_{\rm FS}=92$} fissioning system with a single-Gaussian fragment-yield distribution, that is, \mbox{$\Delta G=0$}. The lower half shows the difference between $\delta_z^{\rm T*}$ and $\delta_z$ (solid black line), and the second derivatives of $\delta_z$ and $\delta_z^{\rm T*}$ multiplied by~3/8 (dashed black and red lines, respectively). In both panels, the red squares are spline interpolations.}
\label{fig4}
\end{figure}

\subsection{The case of $\Delta G=0$}
\label{sec_DG0}

In those cases where $Y^s({\rm Z})$ follows a Gaussian function, \mbox{$\Delta G=0$} and the main source of error in $\delta_z^{\rm T*}$ comes from the approximated average $\langle\langle\delta_z\rangle\rangle$ described in Eqs.~\ref{eq_tracycomps} and~\ref{eq_avdelta}. It is possible to demonstrate that, for small $\delta_z$ (see~\ref{app1}),
\begin{equation}
\langle\langle\delta_z\rangle\rangle\underset{\delta_z\to 0}{\approx}(-1)^Z\left(\frac{3}{8}\partial^2\delta_z+\delta_z\right),
\label{eq_DG0}
\end{equation}
where $\partial^2\delta_z$ is the second derivative with respect to Z, $(\partial^2/\partial{\rm Z}^2)\delta_z$.  The difference between $\delta_z$ and $\delta_z^{\rm T*}$ are then proportional to the second derivative as
\begin{equation}
\frac{3}{8}\partial^2\delta_z\approx\delta_z^{\rm T*}-\delta_z.
\label{eq_difG0}
\end{equation}

This demonstrates that the main source of the difference between $\delta_z^{\rm T*}$ and $\delta_z$ are sharp features in the distribution. A smooth evolution, close to linearity, would not affect the estimation of $\delta_z$ made by $\langle\langle\delta_z\rangle\rangle$.

Figure~\ref{fig4} shows an example of the correlation between the error \mbox{$\varepsilon^{\rm T*}=(\delta_z^{\rm T*}-\delta_z)$} and the second derivative $\partial^2\delta_z$. Indeed, $\partial^2\delta_z$ follows $\varepsilon^{\rm T*}$ for most of the distribution. However, $\partial^2\delta_z$ is not an experimental observable. The use of $\partial^2\delta_z^{\rm T*}$ instead shows a worse performance when following the differences between $\delta_z^{\rm T*}$ and $\delta_z$, particularly around the sharp peak at \mbox{${\rm Z}\approx50$}. This is mainly due to the difference between $\partial^2\delta_z^{\rm T*}$ and $\partial^2\delta_z$: since $\delta_z^{\rm T*}$ is an average of four consecutive $\delta_z$, its value, and consequently its derivatives, are a smoother version of those of $\delta_z$.\footnote{In addition, Eqs.~\ref{eq_DG0} and~\ref{eq_difG0} lose their validity for strong variations of the second derivative within the four-Z window (see Appendix~\ref{app1}).}

\subsection{Impact of random fluctuations}
\label{sec_stat}

As discussed in Sec.~\ref{sec_tracy}, the Log-third-difference formula includes $\langle\langle\delta_z\rangle\rangle$, a weighted average of the even-odd staggering. In the case of an experimental measurement, where a certain degree of statistical uncertainty is not unexpected, the formula cannot distinguish between a true $\delta_z$ staggering and statistical fluctuations. Consequently, in the presence of random statistical fluctuations and \mbox{$\delta_z=0$}, the Log-third-difference formula yields a relatively smooth distribution of these fluctuations that may be interpreted as an actual $\delta_z$, even in the ideal case where \mbox{$\Delta G=0$}.\footnote{In fact, the correlation coefficient of two consecutive $\delta_z^{\rm T}$ varies between 0.3 and~0.5, depending on the relative yields; far from being completely independent.}

\begin{figure}[t]
 \includegraphics[width=\columnwidth]{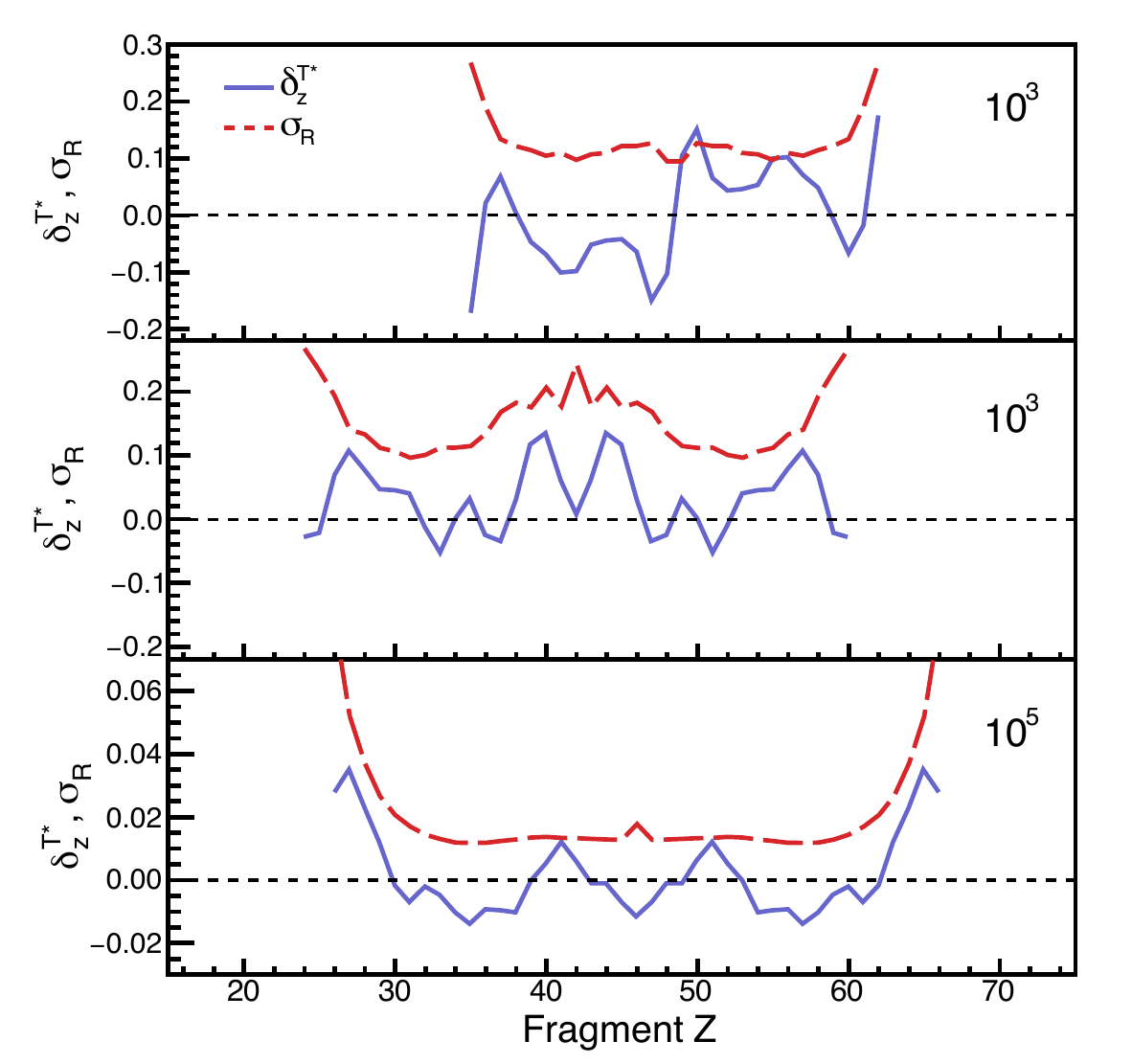}
  \caption{The panels show examples of $\delta_z^{\rm T*}$ calculated for simulated distributions of \mbox{Z$_{\rm FS}=98$, 84, 92} (top, middle, and bottom panel respectively), with \mbox{$\delta_z=0$} and statistical fluctuations corresponding to total yields of~$10^3$ fission events (top and middle panels) and~$10^5$ fission events (bottom panel). The blue solid line is $\delta_z^{\rm T*}$ and the dashed red line is the relative uncertainty $\sigma_R$. Notice the change of scale in the bottom panel.}
\label{fig6}
\end{figure}

Figure~\ref{fig6} shows some examples of this behaviour. Contrary to what is observed with the $\delta_z^{\rm T*}$ obtained for \mbox{$\delta_z=0$} and an underlying sum of Gaussian distributions in Fig.~\ref{fig2}, the $\delta_z^{\rm T*}$ obtained from random statistical fluctuations does not oscillate between positive and negative values and the features of its behaviour may resemble those with a physical origin.

\subsection{The general case}
\label{sec_general}

In a more realistic case, fragment yield distributions are not expected to follow a single Gaussian function nor $\delta_z$ is expected to be constant, but to evolve with the fragment Z and with the possible presence of sharp features like a peak around \mbox{${\rm Z}=50$}~\cite{ramPRC23}. Figure~\ref{fig7} shows the general performance of the Log-third-difference formula when applied to a large collection of simulated distributions as described in the previous section.

The Log-third-difference formula shows a remarkable precision and accuracy, with 95\% of the cases having an error well below 0.01, and an average $\varepsilon^{\rm T*}$ of the order of 10$^{-3}$ for most of the distribution except close to the peak around \mbox{${\rm Z}=50$}. Here, the formula underestimates $\delta_z$, in average, in more than~$-0.05$, and its precision is worse than~0.1 at the centre of the peak. 

Interestingly, around the peak, the precision of $\delta_z^{\rm T*}$ changes its sign and it overestimates the actual value with an average $\varepsilon^{\rm T*}$ of~0.02. This change of character in the accuracy is a consequence of the $\langle\langle\delta_z\rangle\rangle$ component of the formula.

These results reveal a high performance of the Log-third-difference formula for smooth distributions of $\delta_z$ that clearly deteriorates in the presence of sharp features. In the following section, a set of corrections intended to improve the performance of the Log-third-difference formula for such features are proposed.

\section{Corrections to the Log-third-difference formula}
\label{sec_corr}
\subsection{Correction with $\partial^2\delta_z^{\rm T*}$}
\label{sec_corr_d}

Section~\ref{sec_DG0} demonstrates in Eq.~\ref{eq_DG0} a direct relation between the error of the \mbox{even-odd} effect derived from the Log-third-difference formula, $\delta_z^{\rm T*}$, and the second derivative of the even-odd effect amplitude with respect to the fragment Z, $\partial^2\delta_z$. At the same time, the fact that $\partial^2\delta_z$ is not an observable is a limitation to this correction that can be potentially solved by using $\partial^2\delta_z^{\rm T*}$ instead, and defining a corrected $\delta_z^{\rm T*,corr}$ as

\begin{equation}
\delta_z^{\rm T*,corr}=\delta_z^{\rm T*}-\frac{3}{8}\partial^2\delta_z^{\rm T*}
\label{eq_corr}
\end{equation}

\begin{figure}[t]
 \includegraphics[width=\columnwidth]{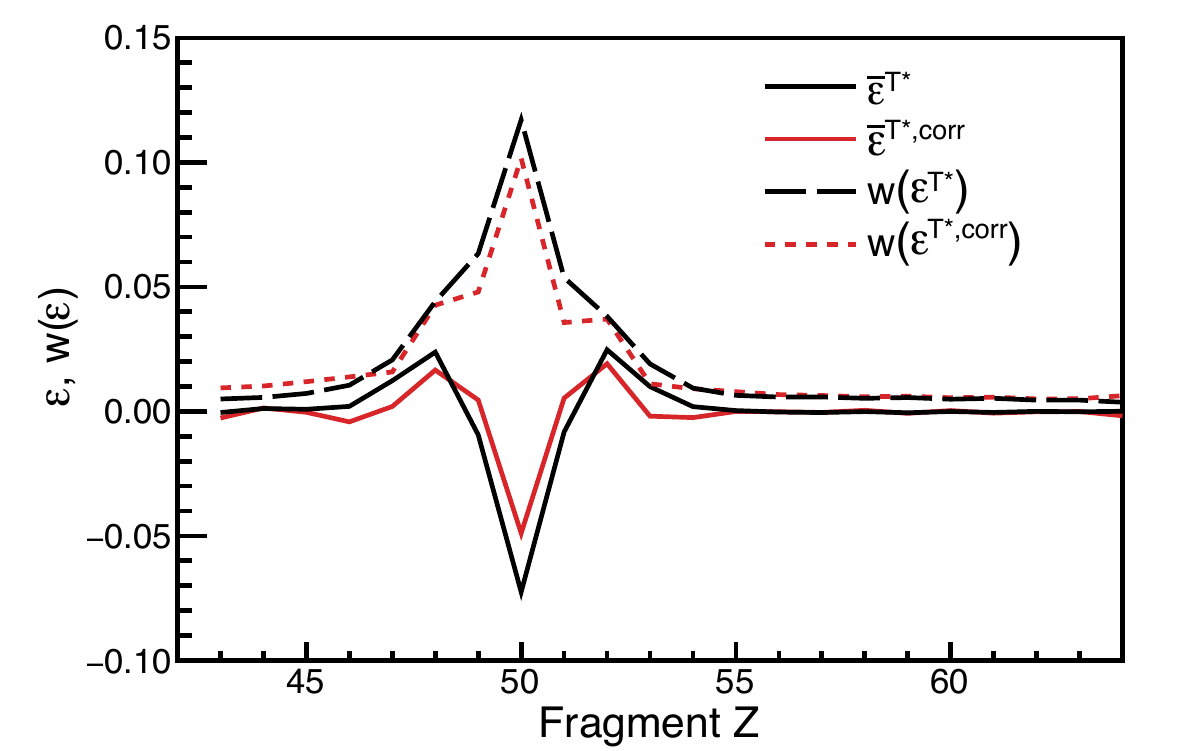}
  \caption{The figure shows the average of $\varepsilon^{\rm T*}$ (black solid line and dots) and $\varepsilon^{\rm T*,corr}$ (red solid line and squares), as a function of the fragment Z for a set of simulated distributions. The black long-dashed and red short-dashed lines show the corresponding $w(\varepsilon^{\rm T*})$ and $w(\varepsilon^{\rm T*,corr})$ respectively.}
\label{fig7}
\end{figure}

Figure~\ref{fig7} shows the average of the error of $\delta_z^{\rm T*}$ compared to the average of the equivalent error of $\delta_z^{\rm T*,corr}$ for a large set of fragment distributions simulated as described in Sec.~\ref{sec_sim}. The figure also shows the corresponding $w$, which encompasses 95\% of the simulated events.

The correction with $\partial^2\delta_z^{\rm T*}$ yields a very modest improvement around the peak at \mbox{${\rm Z}\approx50$}, both in terms of accuracy and precision. This is mainly due to the differences between $\partial^2\delta_z$ and $\partial^2\delta_z^{\rm T*}$, caused by the smoothing effect of the weighted average $\langle\langle\delta_z\rangle\rangle$, as it was previously discussed.

\subsection{Iterative correction}
\label{it_sec}

A key assumption of the calculation of $\delta_z$ is the existence of an underlying $Y^{s}(\rm Z)$ smooth distribution. Therefore, if a $\delta_z$ is obtained by any means, $Y^{s}(\rm Z)$ can be recovered from the observed yields $Y(\rm Z)$ simply by:
\begin{equation}
Y^{s}({\rm Z})=\frac{Y({\rm Z})}{1+(-1)^{\rm Z}\delta_z}.
\label{eq_ys}
\end{equation}

This distribution should have, by definition, no even-odd staggering. Therefore, applying the Log-third-difference formula to $Y^{s}({\rm Z})$ should yield results equivalent to those described in \ref{sec_d0}, that is, compatible with \mbox{$\delta_z=0$}. However, as the computed $\delta_z^{\rm T*}$ is a weighted average of neighbouring $\delta_z$, the yields resulting from applying Eq.~\ref{eq_ys} with $\delta_z^{\rm T*}$ would display a residual even-odd effect resulting from the difference between $\delta_z^{\rm T*}$ and $\delta_z$. But, since the computation of this residual even-odd effect is computed also with the Log-third-difference formula, the result is again a smoothed out distribution of the actual difference.

The correction presented in this section assumes that the residual even-odd effect is smaller than the original $\delta_z^{\rm T*}$, and evaluates the correction in an iterative way. In the first iteration, a tentative smooth yield distribution $Y^{s(0)}({\rm Z})$ is obtained from $\delta_z^{\rm T*}$, denoted $\delta_z^{{\rm T*}(0)}$, as the first step of the iteration:
\begin{equation}
Y^{s(0)}({\rm Z})=\frac{Y({\rm Z})}{1+(-1)^{\rm Z}\delta_z^{{\rm T*}(0)}}.
\label{eq_ys0}
\end{equation}
From this distribution, the residual even-odd effect $\delta_z^{res(0)}$ is calculated with the Log-third-difference formula and Eq.~\ref{eq_tracycorr}. 

A new even-odd effect $\delta_z^{{\rm T*}(1)}$ is obtained from these residuals and the original $\delta_z^{{\rm T*}(0)}$ (see Appendix~\ref{app2}):
\begin{equation}
\delta_z^{{\rm T*}(1)}=\delta_z^{{\rm T*}(0)}+\delta_z^{res(0)}+(-1)^{\rm Z}\delta_z^{{\rm T*(0)}}\delta_z^{res(0)}
\label{eq_d1}
\end{equation}

Further iterations repeat the same process: a provisional smooth yield is obtained in each iteration $i$,
\begin{equation}
Y^{s(i)}({\rm Z})=\frac{Y({\rm Z})}{1+(-1)^{\rm Z}\delta_z^{{\rm T*}(i)}};
\label{eq_ysi}
\end{equation}
the residual even-odd effect $\delta_z^{res(i)}$ of the provisional yields $Y^{s(i)}({\rm Z})$ is calculated; and the resulting $\delta_z^{{\rm T*}(i+1)}$ is
\begin{equation}
\delta_z^{{\rm T*}(i+1)}=\delta_z^{{\rm T*}(i)}+\delta_z^{res(i)}+(-1)^Z\delta_z^{{\rm T*}(i)}\delta_z^{res(i)}.
\label{eq_di}
\end{equation}

As mentioned earlier, the Log-third-difference formula computes the even-odd effect for sets of four Z units, resulting in the average of the effect at a semi-integer value (see Eq.~\ref{eq_tracy}). In order to remove the effect at each Z unit, an interpolation between \mbox{$\delta_z^{{\rm T*}(i)}({\rm Z}-0.5)$} and \mbox{$\delta_z^{{\rm T*}(i)}({\rm Z}+0.5)$} must be performed. In the present case, a cubic spline is used.

\begin{figure}[t]
 \includegraphics[width=\columnwidth]{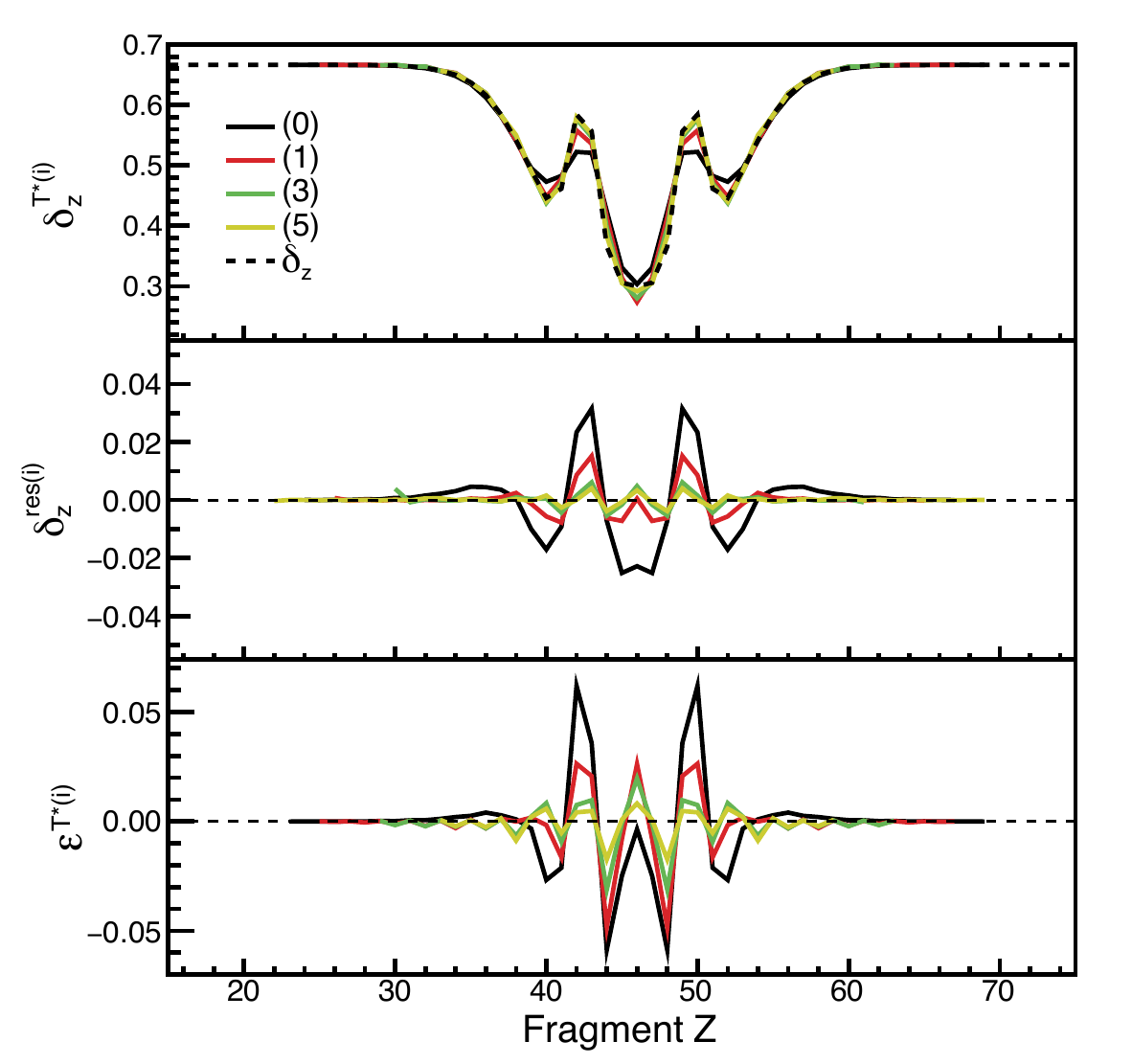}
  \caption{The upper panel shows $\delta_z^{{\rm T*}(i)}$ for the \mbox{$i=0$, 1, 3, and 5} iteration (black, red, green, and yellow solid lines, respectively) for a simulated distribution of a \mbox{Z$_{\rm FS}=96$} fissioning system. The black dashed line is the real $\delta_z$ distribution. The middle and bottom panels show the corresponding $\delta_z^{res(i)}$ and the error $\varepsilon^{{\rm T}*(i)}$ for the same iterations and with the same colour code.}
\label{fig8}
\end{figure}

Figure~\ref{fig8} shows an example of the iterative process. The successive iterations have a clear effect around the peak at \mbox{${\rm Z}\approx 50$}: the residual $\delta_z^{res(i)}$ and the error $\varepsilon^{{\rm T}*(i)}$ distributions diminish with each iteration as $\delta_z^{{\rm T*}(i)}$ approach the real value of $\delta_z$. 

Beyond the region around \mbox{${\rm Z}=50$}, the residual $\delta_z^{res(i)}$ and the error $\varepsilon^{{\rm T}*(i)}$ seem to remain relatively stable although with noticeable oscillations. These are produced by the deviations $\Delta G$ of the underlying yield distribution from a pure Gaussian behaviour discussed in Sec.~\ref{sec_d0}. Since they result from a limitation in the Log-third-difference formula, they do not disappear through any following iteration.\footnote{Even if the actual residual effect is zero, the Log-third-difference formula would still result in \mbox{$\Delta G \neq 0$}, unless the underlying distribution is a single Gaussian.} Interestingly, the observation of these persistent oscillations signals the end of the net effect of the iterations, which, for the smooth regions of $\delta_z$, it can be as soon as after the first iteration.

The effect of this iterative correction is very evident in sharp features as peaks or sudden changes that are smoothed out by the $\langle\langle\delta_z\rangle\rangle$ component of the Log-third-difference formula, and that are hardly recovered by the second-derivative correction discussed in Sec.~\ref{sec_corr_d}.

\begin{figure}[t]
 \includegraphics[width=\columnwidth]{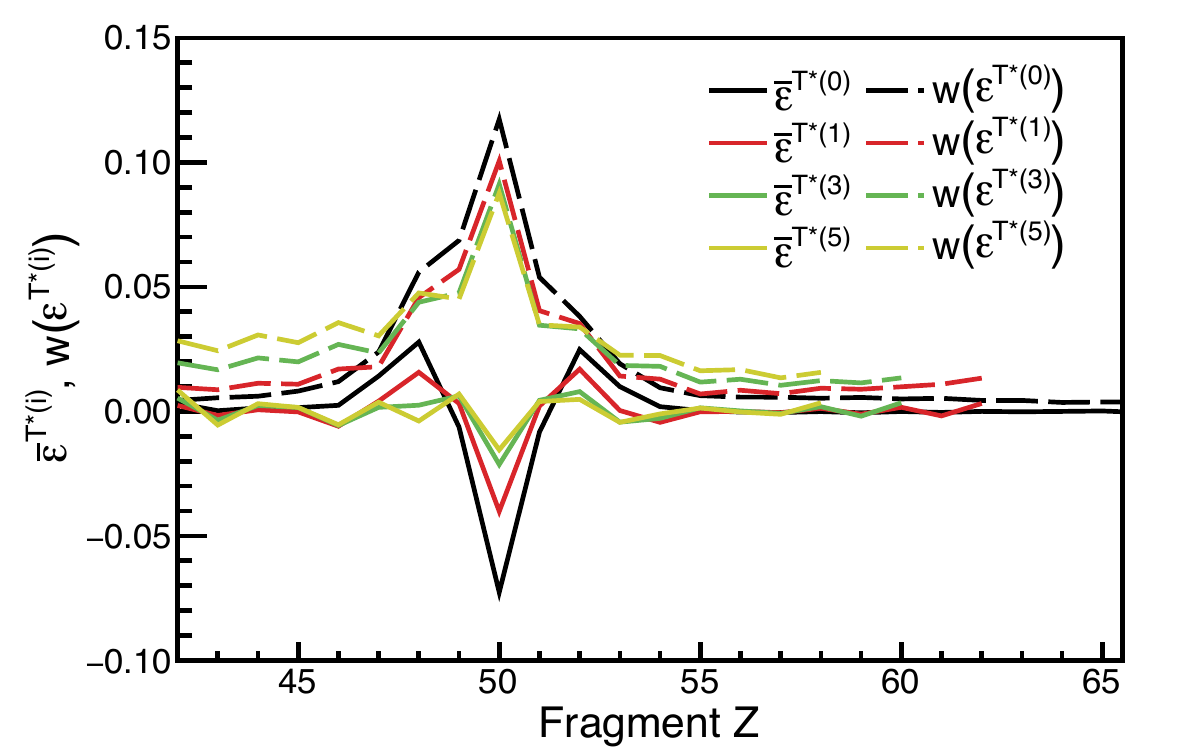}
  \caption{The figure shows the average of the error $\varepsilon^{{\rm T*}(i)}$ (solid lines) and the corresponding $w(\varepsilon^{{\rm T*}(i)})$ (dashed lines) for the \mbox{$i=0$, 1, 3, and 5} iteration (black, red, green, and yellow solid lines, respectively) as a function of the fragment Z for a set of simulated distributions.}
\label{fig9}
\end{figure}

Figure~\ref{fig9} shows the effect of the iterative process in both the average error $\bar\varepsilon^{{\rm T}*(i)}$ and its width $w(\varepsilon^{{\rm T}*(i)}$). Concerning the average $\bar\varepsilon^{{\rm T}*(i)}$, there is a clear improvement around \mbox{${\rm Z}=50$} with the subsequent iterations: The amplitude of $\bar\varepsilon^{{\rm T}*(i)}$ is reduced in more than a factor seven, from more than~$-0.07$ down to~$-0.01$ in the \mbox{$i=3$} iteration. The width is also reduced and 95\% of the cases are within \mbox{$\varepsilon^{{\rm T}*(i)} = \pm 0.1$} already in the first iteration. 

Outside of the influence of the \mbox{${\rm Z}\approx50$} region, the iterative process brings the aforementioned oscillations associated with $\Delta G$. It is interesting to note that, outside the peak region where a smooth behaviour of $\delta_z$ is expected, the \mbox{$i=0$} iteration, that is the application of the Log-third-difference formula, seems to fare better than the next ones. This is particularly true when looking at the width $w(\varepsilon^{{\rm T}*(i)})$ of the distributions: each subsequent iteration seems to increase the width, except around the peak at \mbox{${\rm Z}=50$}. 

An unwanted limitation of the iterative process is that the region of the distribution corrected in each iteration is smaller than the previous one. Since each iteration involves the application of the Log-third-difference formula to compute $\delta_z^{res(i)}$, the limits in Z are reduced in two units each. However, in most of the systems of interest, the sharp features appear around \mbox{${\rm Z}=50$}, which is typically far from the tails of the fragment distributions, thus several steps of the iterative process can be applied.

\subsection{Multi-Gaussian fit correction}
\label{gaus_sec}

The iterative correction of \ref{it_sec} tries to minimise the effect of the $\langle\langle\delta_z\rangle\rangle$ component in the Log-third-difference formula with the iterative removal of the remaining even-odd staggering. In another strategy, these remaining features can be minimised by assuming that the underlying smooth yield distribution can be approximated by a multi-Gaussian function.

In the first step, a smooth $Y^{s(0)}({\rm Z})$ distribution, such as the one computed in Eq.~\ref{eq_ys0}, is calculated as starting point. The second step consists in fitting this distribution to the expected function that describes the actual $Y^{s}({\rm Z})$. Usually, this would be a multi-Gaussian function with each Gaussian distribution representing a fission yield mode, similar to the one shown in Figs.~\ref{fig1}~and~\ref{fig2}. In the third and final step, the amplitude of the even-odd effect is obtained from the measured yields $Y({\rm Z})$ and the fitted multi-Gaussian function $F^{mG}({\rm Z})$:
\begin{equation}
\delta_z^{{\rm G}}=(-1)^Z\left(\frac{Y({\rm Z})}{F^{mG}({\rm Z})}-1\right).
\label{eq_fitG}
\end{equation}

\begin{figure}[t]
 \includegraphics[width=\columnwidth]{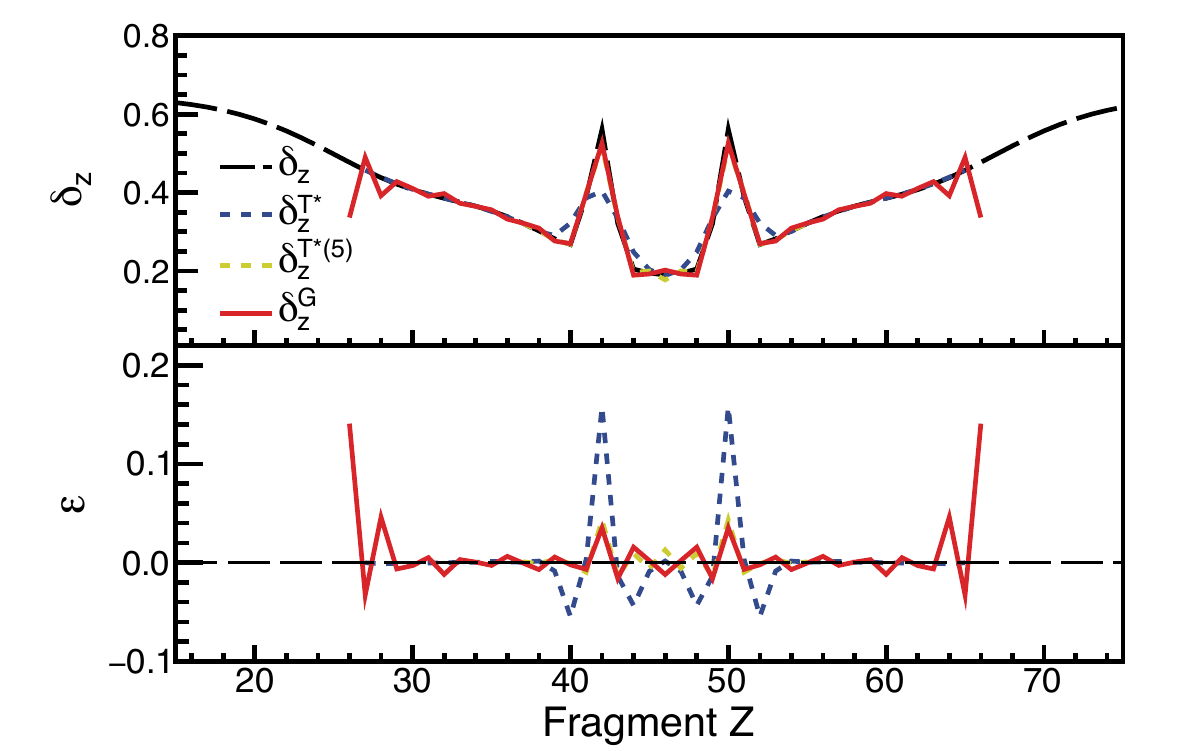}
  \caption{The upper half of the figure shows a simulated $\delta_z$ (long-dashed black line) with the resulting $\delta_z^{\rm T*}$ (dashed blue line), $\delta_z^{{\rm T*}(5)}$ (dashed yellow line), and $\delta_z^{\rm G}$ (red line) for a \mbox{Z$_{\rm FS} = 90$} fissioning system. The lower half shows, with the same colour code, the error $\varepsilon$ associated with each estimation of $\delta_z$. Notice that the high similarity between $\delta_z^{{\rm T*}(5)}$ and $\delta_z^{\rm G}$ makes it somehow difficult to tell their respective lines apart.}
\label{fig10}
\end{figure}

Figure~\ref{fig10} shows the result of a typical application of this process compared to those of the Log-third-difference formula and the iterative correction described in~\ref{it_sec}. The distribution of $\delta_z^{\rm G}$ is able to recover the features of the peak at \mbox{${\rm Z}\approx50$}, which are partially lost to the Log-third-difference formula, with a similar performance as the fifth step of iterative process $\delta_z^{{\rm T*}(5)}$. However, also similar to the iterative process, clear oscillations appear along the whole distribution, although, in this case, the origin is different.

As explained before, the fit of the initial $Y^{s(0)}({\rm Z})$ results in a multi-Gaussian function $F^{mG}({\rm Z})$. In most of the cases, this function is (hopefully) slightly different from the actual $Y^{s}({\rm Z})$ in a quantity than can be defined as a function of Z as \mbox{$\varepsilon^{mG}({\rm Z})=F^{mG}({\rm Z})-Y^{s}({\rm Z})$}. It can be easily shown that $\delta_z^{\rm G}$ depends on this difference as
\begin{equation}
\delta_z^{{\rm G}}=\delta_z\left(1-\frac{\varepsilon^{mG}({\rm Z})}{F^{mG}({\rm Z})}\right)-(-1)^{\rm Z}\frac{\varepsilon^{mG}({\rm Z})}{F^{mG}({\rm Z})}.
\label{eq_errmG}
\end{equation}
That is, in the regions of Z where there is a difference between the fitted function and $Y^{s}({\rm Z})$, the resulting $\delta_z^{{\rm G}}$ is scaled by the relative error of $F^{mG}({\rm Z})$ and oscillates with an amplitude that is proportional to the relative error $\varepsilon^{mG}/F^{mG}$ and with a period of one Z unit.

\begin{figure}[t]
 \includegraphics[width=\columnwidth]{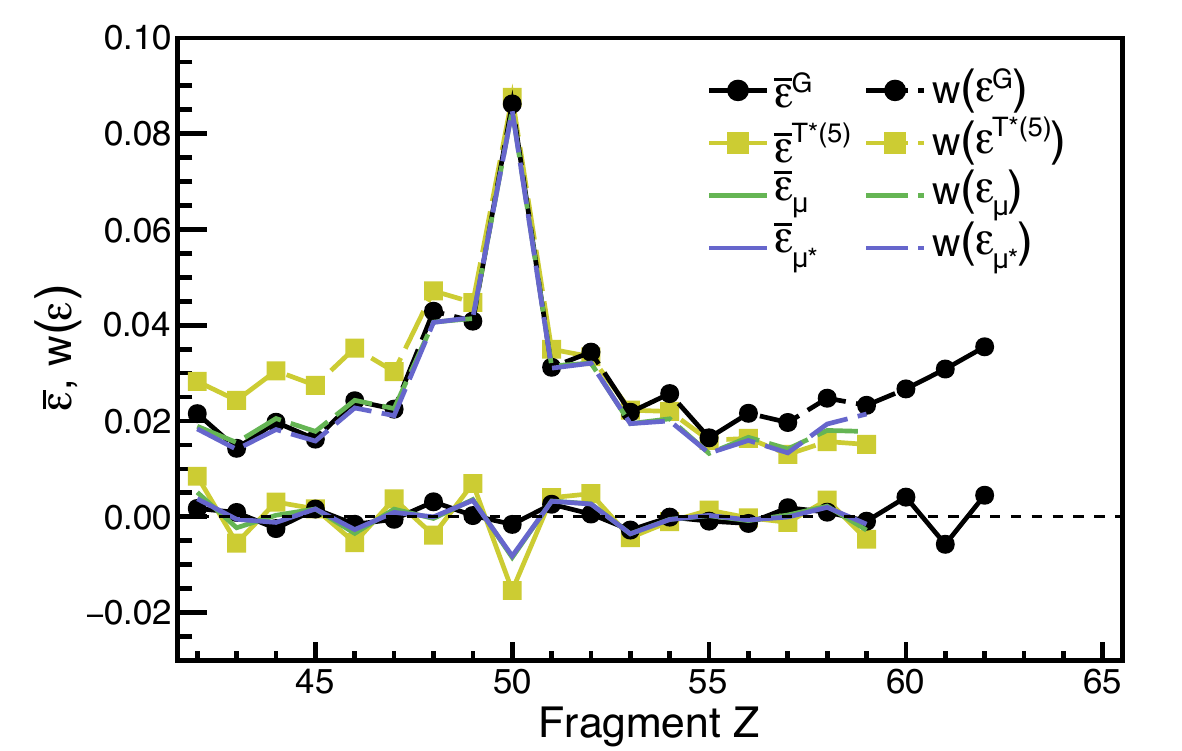}
  \caption{The figure shows the average of the error $\delta_z^{\rm G}$, ${\bar{\varepsilon}}^{\rm G}$, and the corresponding $w(\varepsilon^{\rm G})$ (solid and dashed black lines and dots, respectively), compared to $\bar{\varepsilon}^{{\rm T*}(5)}$ and $w(\varepsilon^{{\rm T*}(5)})$ (solid and dashed yellow lines and squares, respectively) as a function of the fragment Z for a set of simulated distributions. The solid and dashed lines correspond to the average error and width of a simple average of $\delta_z^{\rm G}$ and $\delta_z^{{\rm T*}(5)}$ (green) and an average taking into account the correlation between both (blue).}
\label{fig11}
\end{figure}

Figure~\ref{fig11} shows the average error and width of $\delta_z^{{\rm G}}$ for a set of simulated distributions, compared to those of $\delta_z^{\rm T*(5)}$ from the iterative correction. In the figure, $\delta_z^{\rm G}$ fares better in average than $\delta_p^{{\rm T*}(5)}$, with smaller oscillations and more accuracy around \mbox{${\rm Z}=50$}. The width of the distribution is noticeably smaller for $\delta_z^{\rm G}$ close to the symmetry. However, the precision degrades quickly towards high asymmetry and lower yields. Around \mbox{${\rm Z}=50$}, both approaches show the same precision, with 95\% of the cases below an error of 0.1.

In any case, it should be noted that the examples and data shown here are obtained from relatively simple fits. That is, in order to compute $\delta_z^{\rm G}$ for the collection of simulated distributions in a reasonable time, each distribution was fitted in a single step, through $\chi^2$ minimisation, to a multi-Gaussian function with two asymmetric and one symmetric modes. More dedicated and careful fits can potentially yield better results.

\subsection{Correlation between $\delta_z^{\rm G}$ and $\delta_z^{{\rm T*}(i)}$}

In principle, since they are calculated with different approaches, $\delta_z^{\rm G}$ and $\delta_z^{{\rm T*}(i)}$ can be considered as being independent estimators of $\delta_z$. Thus, one might be tempted to improve the calculation with some sort of average.

\begin{figure}[t]
 \includegraphics[width=\columnwidth]{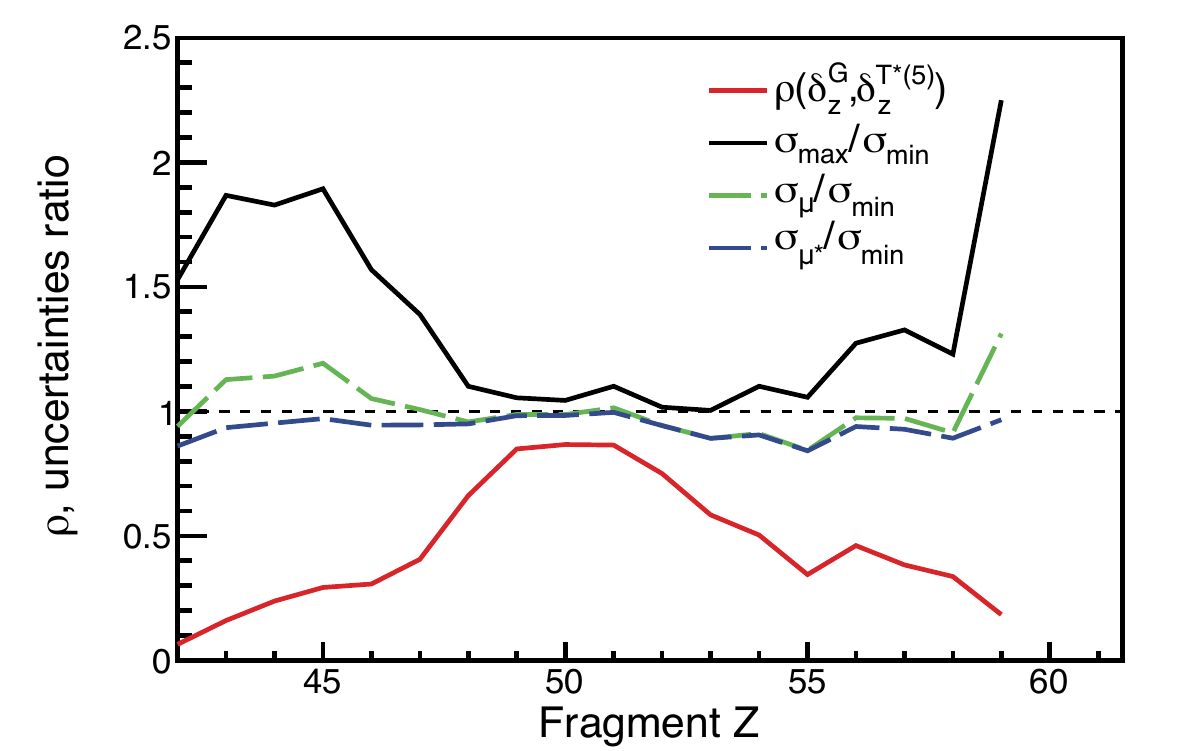}
  \caption{The figure shows the correlation coefficient $\rho$ between $\delta_z^{\rm G}$ and $\delta_z^{{\rm T*}(5)}$, and the ratio between their uncertainties as a function of the fragment Z (solid red and black lines respectively). The dashed lines are the uncertainty of a simple average, $\sigma_{\mu}$ (green), and that of an average taking correlations into account, $\sigma_{\mu*}$ (blue), both normalised to the smallest of $\sigma_{\delta_z^{\rm G}}$ and $\sigma_{\delta_z^{{\rm T*}(5)}}$. Values below~1 show a potential improvement in the uncertainty of the corresponding average.}
\label{fig12}
\end{figure}

However, both share the same initial step and begin with a calculation of $\delta_z^{\rm T*}$ with the Log-third-difference formula. This has a certain effect on their correlation, more in particular around \mbox{${\rm Z}=50$}, as it is shown with the correlation coefficient $\rho$ in Fig.~\ref{fig12}. There is a significant variation between the values around ${\rm Z}=50$, where $\rho$ reaches 0.9, and  close to symmetry, where it goes down to almost~0.1. For more asymmetric splits, beyond \mbox{${\rm Z}=54$}, $\rho$ fluctuates around~0.4.

Any correlation between $\delta_z^{\rm G}$ and $\delta_z^{{\rm T*}(5)}$ reduces the potential gain of performing an average. It can be shown that the improvement when performing a simple average depends on $\rho$ and the ratio of both uncertainties: there is a net improvement only when the ratio of uncertainties is smaller than \mbox{$\sqrt{\rho^2+3}-\rho$} (see Appendix~\ref{app3}).  Unfortunately, this is rarely the case for $\delta_z^{\rm G}$ and $\delta_z^{{\rm T*}(5)}$. As Figure~\ref{fig12} shows, the precision of the average value is hardly an improvement around the peak and, moreover, it degrades slightly the precision obtained by the best of $\delta_z^{\rm G}$ and $\delta_z^{{\rm T*}(5)}$ alone.

A more sophisticated average can include the effect of these correlations and, contrary to a simple average, the uncertainty of the result would be always better than the best of both $\delta_z^{\rm G}$ and $\delta_z^{{\rm T*}(5)}$ (see Appendix~\ref{app3}). However, even in this case, the improvement is very modest, as Fig.~\ref{fig12} demonstrates, and it hardly reaches more than 10\% better. Moreover, around the \mbox{${\rm Z}=50$} peak, there is barely any improvement at all. Figure~\ref{fig11} shows the results in terms of $w$ and $\bar{\varepsilon}$ of computing a simple average of $\delta_z^{\rm G}$ and $\delta_z^{{\rm T*}(5)}$ or by taking into account the correlation between both.

Beyond the correlation between $\delta_z^{\rm G}$ and $\delta_z^{{\rm T*}(5)}$, the systematic underestimation of $\delta_z^{{\rm T*}(i)}$ also advises against any form of average. The error of $\delta_z^{\rm G}$ appears more stable and closer to 0, therefore any average with a positive correlated quantity with a similar uncertainty, as it is the case of $\delta_z^{{\rm T*}(5)}$, would result in a biased average, with or without taking into account correlations. This can be easily observed in Fig.~\ref{fig11}: any average between $\delta_z^{\rm G}$ and $\delta_z^{{\rm T*}(5)}$ result in an error around \mbox{${\rm Z}=50$} that underestimates the real value of $\delta_z(50)$.

\section{Summary and conclusions}

The even-odd staggering in fission fragments is an important observable for the study of the role that intrinsic excitations play in the fission process. As such, one can define the even-odd staggering in neutron or proton number, although the most accesible is the proton even-odd staggering. The proton even-odd effect is traditionally described as the modulation of an underlying smooth distribution of the measured elemental fission yields. 

Among the several ways to estimate the evolution of its amplitude as a function of the fragment proton number, the Log-third-difference method, first proposed by Tracy et al.~\cite{tracy}, computes the amplitude in a four-Z window by assuming an underlying Gaussian behaviour and a constant even-odd effect. In this formula, two components contribute to the result: the difference between the Gaussian behaviour and the real one; and the average of the even-odd effect distribution in the window. However, measured even-odd effects present sharp features, particularly around \mbox{${\rm Z}=50$}, that are a challenge for the Log-third-difference formula, mostly due to the weighted average that smoothes out the distribution.

Three strategies to diminish the error of the formula in these sharp regions are presented in this work: the use of the second-derivative of the even-odd effect, an iterative application of the formula, and a multi-Gaussian fit. The use of the second derivative gives very modest results. The iterative correction improves in a very significant way the estimation of the sharp features. However, each iteration restricts the range of application, and a residual underestimation of the effect at the maximum of a sharp peak remains. The use of a multi-Gaussian fit has a noticeable better performance than the iterative process. While it degrades towards the tails of the distribution at high asymmetry, the multi-Gaussian fit shows the best accuracy of the three methods presented. However, it is worthy to recall that even though the precision and accuracy improve around sharp features in the iterative and multi-Gaussian fit methods, it is not better than the direct application of the Log-third-difference formula in smooth regions.

In conclusion, both the iterative and multi-Gaussian fit methods offer a considerable improvement in precision and accuracy in the calculation of an even-odd effect distribution with sharp features as the observed peaks at \mbox{${\rm Z}\approx50$}. This can be particularly useful in studies that treat the amplitude, and not only the evolution, of the even-odd effect as a relevant observable.

\backmatter

\bmhead{Acknowledgements}

M. C. acknowledges the financial support from the Xunta de Galicia (CIGUS Network of Research Centres) and the European Union, the financial support from the Spanish Agencial Estatal de Investigaci\'on under grant PGC2018-096717-B-C22, and the support by María de Maeztu grant CEX2023-001318-M funded by MICIU/AEI/10.13039/501100011033. Part of this work is included in the End-of-Degree project of Bel\'en Montenegro Vi\~nas presented at the Universidade de Santiago de Compostela in July, 2023.

\begin{appendices}

\section{Derivation of some of the equations in the text}
\subsection{Equation~\ref{eq_stat}}
\label{app0}

The uncertainty of $\delta_z^{\rm T*}$ can be computed with the classical propagation of errors as
\begin{equation}
\sigma_{\delta_z^{\rm T*}}^2=\sum_i\left(\frac{\partial\delta_z^{\rm T*}}{\partial Y(\rm Z_i)}\right)^2\sigma_{i}^2.
\end{equation}
The partial derivative can be factorised considering Eqs.~\ref{eq_tracy} and \ref{eq_tracycorr} as
\begin{equation}
\sigma_{\delta_z^{\rm T*}}^2=\sum_i\left(\frac{\partial\delta_z^{\rm T*}}{\partial \delta_z^{\rm T}}\right)^2\left(\frac{\partial\delta_z^{\rm T}}{\partial Y(\rm Z_i)}\right)^2\sigma_{i}^2.
\label{eq_fact}
\end{equation}

The first partial derivative can be obtained from Eqs.~\ref{eq_tracycorr} and \ref{eq_G1}:
\begin{equation}
\frac{\partial\delta_z^{\rm T*}}{\partial \delta_z^{\rm T}}=\frac{4e^{2\delta_z^{\rm T}}}{\left(e^{2\delta_z^{\rm T}}+1\right)^2}=1-\left({\delta_z^{\rm T*}}\right)^2,
\end{equation}
while the second one depends on Z$_{i}$ as
\begin{equation}
\frac{\partial\delta_z^{\rm T}}{\partial Y({\rm Z_{0,3}})}=\frac{1}{8}\frac{1}{Y({\rm Z_{0,3}})};~~\frac{\partial\delta_z^{\rm T}}{\partial Y({\rm Z_{1,2}})}=\frac{3}{8}\frac{1}{Y({\rm Z_{1,2}})}.
\end{equation}

Approximating the statistical uncertainty by a constant fraction of the yields \mbox{$\sigma_R\approx\sigma_i / Y({\rm Z_i})$},
\begin{equation}
\frac{\partial\delta_z^{\rm T}}{\partial Y({\rm Z_{0,3}})}\sigma_{0,3}\approx \frac{1}{8}\sigma_R;~~\frac{\partial\delta_z^{\rm T}}{\partial Y({\rm Z_{1,2}})}\sigma_{1,2}\approx \frac{3}{8}\sigma_R.
\end{equation}

Finally, introducing these in Eq.~\ref{eq_fact} and assuming that \mbox{$\delta_z^{\rm T*}\approx \delta_z$}:
\begin{equation}
\sigma_{\delta_z^{\rm T*}}^2\approx\frac{20}{8^2}\sigma_R^2\left(1-\delta_z^2\right)^2.
\end{equation}

Taking the square root, Eq.~\ref{eq_stat} is obtained.

\subsection{Equation~\ref{eq_DG0}}
\label{app1}

The average $\langle\langle\delta_z\rangle\rangle$ can be expressed in terms of the real $\delta_z({\rm Z})$. In particular, $\langle\langle\delta_z\rangle\rangle$(Z$_{1.5}$) is calculated with the $\delta_z$(Z) values from Z$_{0}$ to Z$_{3}$ as in Eq.~\ref{eq_avdelta}. Considering that \mbox{$\ln(1\pm\delta_z)\approx\pm\delta_z$} when \mbox{$\delta_z\to 0$}, it can be rearranged into 
\begin{equation}
\begin{aligned}
\langle\langle\delta_z\rangle\rangle\approx\frac{(-1)^{Z_0}}{8}\Big[\delta_z({\rm Z_{0}})+3\delta_z({\rm Z_{1}})\\
+3\delta_z({\rm Z_2})+\delta_z({\rm Z_3})\Big].
\end{aligned}
\label{eq_app_2}
\end{equation}

The behaviour of $\delta_z({\rm Z})$ can be described as a Taylor series around Z$_{1.5}$. Up to second order, it would be
\begin{equation}
\begin{aligned}
\delta_z({\rm Z})\approx\delta_z({\rm Z_{1.5}})+(Z-Z_{1.5})\frac{\partial\delta_z}{\partial Z}\Bigg|_{Z_{1.5}}\\
+\frac{(Z-Z_{1.5})^2}{2}\frac{\partial^2\delta_z}{\partial Z^2}\Bigg|_{Z_{1.5}}
\end{aligned}
\label{eq_app_3}
\end{equation}

Each of the terms in Eq.~\ref{eq_app_2}, can be described with this series:
\begin{equation}
\begin{aligned}
\delta_z&({\rm Z_{0}})=\delta_z({\rm Z_{1.5}})-\frac{3}{2}\frac{\partial\delta_z}{\partial Z}\Bigg|_{Z_{1.5}}+\frac{9}{8}\frac{\partial^2\delta_z}{\partial Z^2}\Bigg|_{Z_{1.5}}\\
\delta_z&({\rm Z_{1}})=\delta_z({\rm Z_{1.5}})-\frac{1}{2}\frac{\partial\delta_z}{\partial Z}\Bigg|_{Z_{1.5}}+\frac{1}{8}\frac{\partial^2\delta_z}{\partial Z^2}\Bigg|_{Z_{1.5}}\\
\delta_z&({\rm Z_{2}})=\delta_z({\rm Z_{1.5}})+\frac{1}{2}\frac{\partial\delta_z}{\partial Z}\Bigg|_{Z_{1.5}}+\frac{1}{8}\frac{\partial^2\delta_z}{\partial Z^2}\Bigg|_{Z_{1.5}}\\
\delta_z&({\rm Z_{3}})=\delta_z({\rm Z_{1.5}})+\frac{3}{2}\frac{\partial\delta_z}{\partial Z}\Bigg|_{Z_{1.5}}+\frac{9}{8}\frac{\partial^2\delta_z}{\partial Z^2}\Bigg|_{Z_{1.5}}
\end{aligned}
\label{eq_app_4}
\end{equation}

Plugging these terms in Eq.~\ref{eq_app_2} results in Eq.~\ref{eq_DG0}.

This approximation is correct as long as the second derivative is relatively constant within the four-Z window and derivatives beyond second order are negligible. 

\subsection{Equation~\ref{eq_d1}}
\label{app2}

The basic assumption is that the yield distribution can be described with an underlying smooth distribution $Y^{s}({\rm Z})$ with an even-odd effect on top of it:
\begin{equation}
Y=Y^{s}[1+(-1)^Z\delta_z]
\label{eq_app_5}
\end{equation}
Once the first iteration of even-odd effect $\delta_z^{{\rm T*}(0)}$ is obtained with the Log-third-difference formula, the residual even-odd effect $\delta_z^{res(0)}$ is defined as
\begin{equation}
Y^{s(0)}\left[1+(-1)^Z\delta_z^{res(0)}\right]=\frac{Y}{1+(-1)^Z\delta_z^{{\rm T*}(0)}}.
\label{eq_app_6}
\end{equation}
According to Eq.~\ref{eq_app_5}, the resulting even-odd effect $\delta_z^{{\rm T*}(1)}$ comprising the original $\delta_z^{{\rm T*}(0)}$ and the residual $\delta_z^{res(0)}$, after the first iteration, can be obtained as
\begin{equation}
\begin{aligned}
Y&=Y^{s(0)}\left[1+(-1)^Z\delta_z^{{\rm T*}(1)}\right]\\
Y&=Y^{s(0)}\left[1+(-1)^Z\delta_z^{res(0)}\right]\left[1+(-1)^Z\delta_z^{{\rm T*}(0)}\right],
\end{aligned}
\label{eq_app_7}
\end{equation}
therefore
\begin{equation}
\begin{aligned}
&1+(-1)^Z\delta_z^{{\rm T*}(1)}=\\
&=\left[1+(-1)^Z\delta_z^{res(0)}\right]\left[1+(-1)^Z\delta_z^{{\rm T*}(0)}\right]
\label{eq_app_8}
\end{aligned}
\end{equation}
and, finally,
\begin{equation}
\delta_z^{{\rm T*}(1)}=\delta_z^{res(0)}+\delta_z^{{\rm T*}(0)}+(-1)^Z\delta_z^{res(0)}\delta_z^{{\rm T*}(0)}.
\label{eq_app_9}
\end{equation}

\section{Impact of correlations on the uncertainty of the average}
\label{app3}

In the most simple case, the average of two quantities $a$ and $b$ is:
\begin{equation}
\mu=\frac{1}{2}(a+b).
\end{equation}

The uncertainty of this average is related with their uncertainties, $\sigma_a$ and $\sigma_b$, and the correlation coefficient $\rho$ through
\begin{equation}
\sigma_{\mu}^2=\frac{1}{4}(\sigma_a^2+\sigma_b^2+2\rho\sigma_a\sigma_b).
\label{eq_mu}
\end{equation}

An improvement of the average with respect to the individual values happens when $\sigma_{\mu}$ is smaller than the smallest of $\sigma_a$ and \mbox{$\sigma_b$. If $\sigma_a\leq\sigma_b$}, only when \mbox{$\sigma_{\mu}/\sigma_{a}\leq 1$} it is worthy to take the average.
\begin{figure}[t] 
\includegraphics[width=\columnwidth]{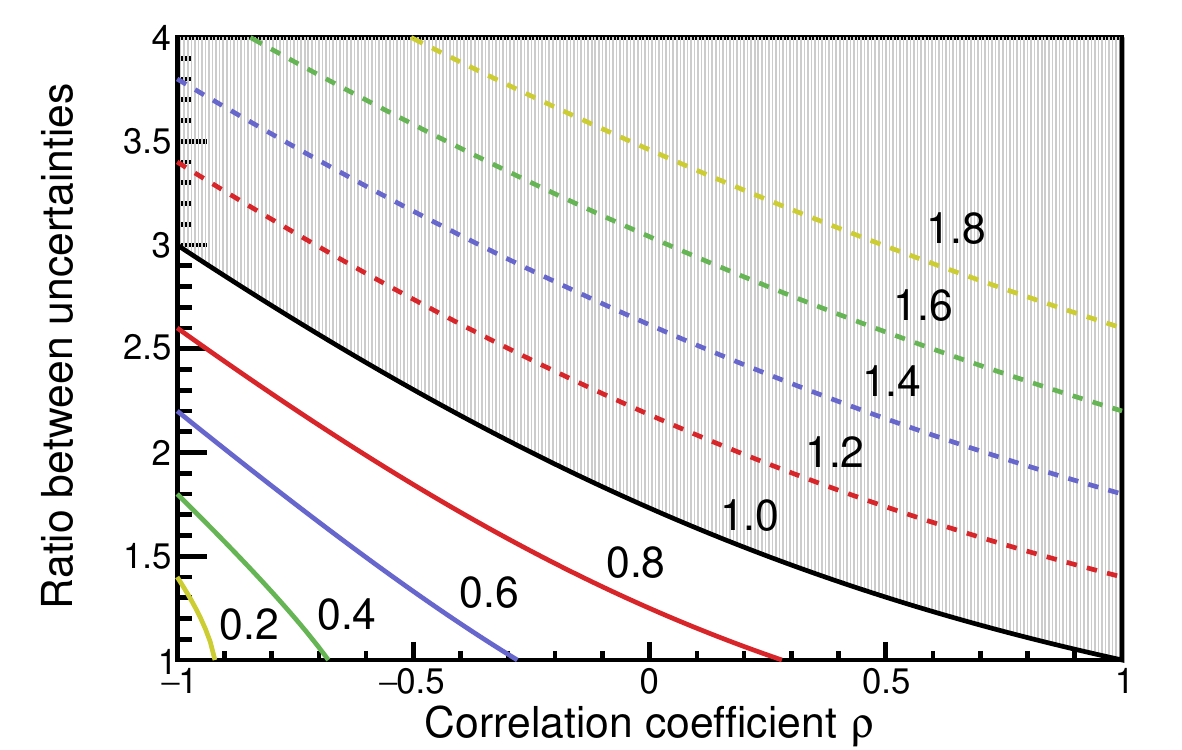}
  \caption{The lines show the relation between the ratio between uncertainties \mbox{$\sigma_b/\sigma_a$} and the correlation coefficient $\rho$. Each colour and number corresponds to a different value of \mbox{$\sigma_{\mu}/\sigma_a$}. The values outside of the shaded area and solid lines would result in a net improvement with \mbox{$\sigma_{\mu}/\sigma_a\leq 1$}. Inside this area, the average would have an uncertainty bigger than $\sigma_a$, thus it would not yield any improvement.}
\label{figapp1}
\end{figure}
\begin{figure}[b]
 \includegraphics[width=\columnwidth]{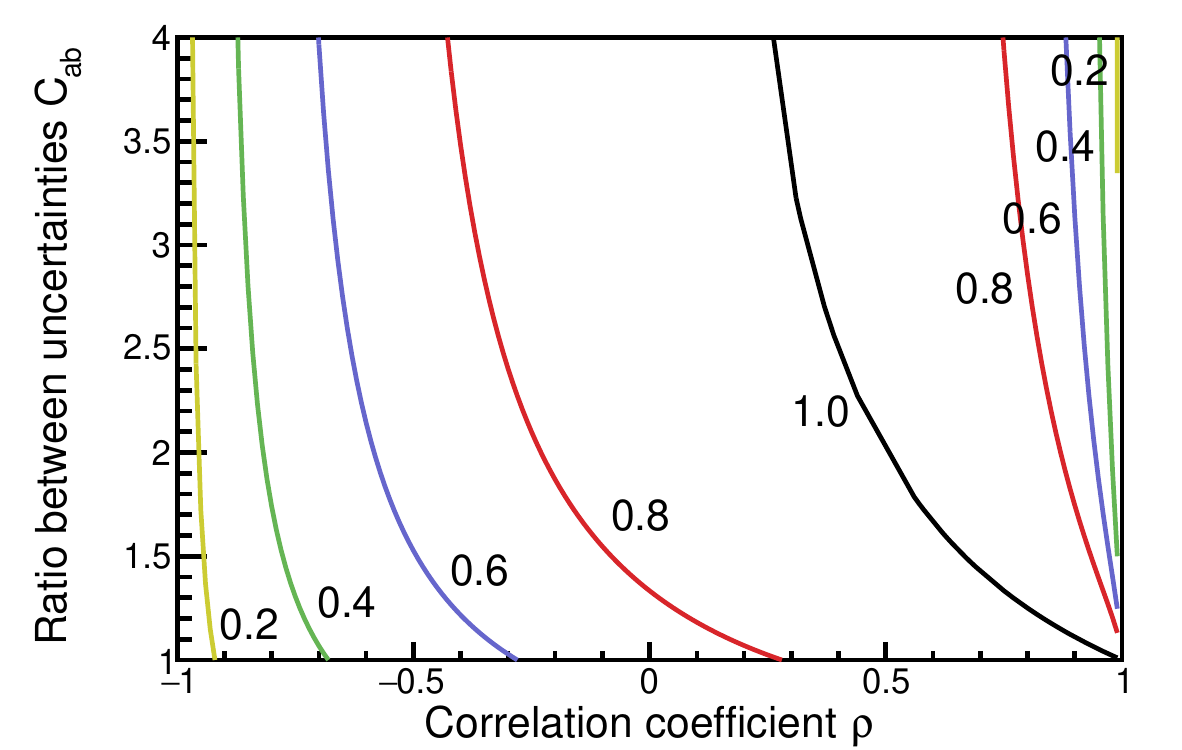}
  \caption{The lines show the relation between the ratio between uncertainties \mbox{$\sigma_b/\sigma_a$} and the correlation coefficient $\rho$. Each colour and number corresponds to different values of \mbox{$\sigma_{\mu*}/\sigma_a$}. Contrary to the previous figure, all values would result in a net improvement with \mbox{$\sigma_{\mu}/\sigma_a\leq 1$}.}
\label{figapp2}
\end{figure}
The value of $\sigma_{\mu}/\sigma_{a}$ only depends on $\rho$ and the ratio between uncertainties $\sigma_b/\sigma_a$, which, in this case, is defined as bigger than 1. Equation~\ref{eq_mu} can be expressed as
\begin{equation}
\left(\frac{\sigma_{\mu}}{\sigma_{a}}\right)^2=\frac{1}{4}\left[1+\left(\frac{\sigma_{b}}{\sigma_{a}}\right)^2+2\rho\left(\frac{\sigma_{b}}{\sigma_{a}}\right)\right],
\end{equation}
and the ratio between uncertainties can be obtained for specific values of \mbox{$\sigma_{\mu}/\sigma_{a}$}:
\begin{equation}
\frac{\sigma_{b}}{\sigma_{a}}=-\rho+\sqrt{\rho^2-1+4\left(\frac{\sigma_{\mu}}{\sigma_{a}}\right)^2}.
\label{eq_crho}
\end{equation}

Figure~\ref{figapp1} shows the relation between \mbox{$\sigma_b/\sigma_a$} and $\rho$ for a series of improvement values. It is easy to see that, in order to obtain a net improvement, \mbox{$\sigma_b/\sigma_a\leq -\rho + \sqrt{\rho^2+3}$}.

A more sophisticated average would take into account the uncertainties and correlation of both values. It can be shown that, following the Gauss-Markov theorem, the average is~\cite{sch95}
\begin{equation}
\mu^*=\frac{\frac{a}{\sigma_a^2}+\frac{b}{\sigma_b^2}-(a+b)\frac{\rho}{\sigma_a\sigma_b}}{\frac{1}{\sigma_a^2}+\frac{1}{\sigma_b^2}-\frac{2\rho}{\sigma_a\sigma_b}},
\end{equation}
and its variance is
\begin{equation}
\sigma_{\mu^*}^2=\frac{(1-\rho^2)}{\frac{1}{\sigma_a^2}+\frac{1}{\sigma_b^2}-\frac{2\rho}{\sigma_a\sigma_b}}.
\end{equation}

In this case, the ratio \mbox{$\sigma_{\mu*}/\sigma_a$} is
\begin{equation}
\left(\frac{\sigma_{\mu*}}{\sigma_{a}}\right)^2=\frac{(1-\rho^2)}{1+\left(\frac{\sigma_b}{\sigma_a}\right)^{-2}-2\rho\left(\frac{\sigma_b}{\sigma_a}\right)^{-1}},
\end{equation}
and the ratio between uncertainties,
\begin{equation}
\begin{aligned}
\frac{\sigma_b}{\sigma_a}&=\frac{\rho}{\left(1-\frac{1-\rho^2}{\left(\frac{\sigma_{\mu*}}{\sigma_a}\right)^2}\right)}\\
&+\sqrt{\frac{\rho^2}{\left(1-\frac{1-\rho^2}{\left(\frac{\sigma_{\mu*}}{\sigma_a}\right)^2}\right)^2}-\frac{1}{\left(1-\frac{1-\rho^2}{\left(\frac{\sigma_{\mu*}}{\sigma_a}\right)^2}\right)}}.
\end{aligned}
\end{equation}

For \mbox{$\rho\in(-1,1)$} and \mbox{$\sigma_b/\sigma_a\in(1,\infty)$}, \mbox{$\sigma_{\mu*}/\sigma_a\leq1$}. That is, performing the average with the correlations taken into account always results in an improved value. Although the level of improvement depends on \mbox{$\sigma_b/\sigma_a$} and $\rho$, as Fig.~\ref{figapp2} shows.

\end{appendices}

\bibliography{sn-bibliography}%

\end{document}